\documentclass[a4paper]{amsart}
\usepackage{amsmath,amssymb,amsthm}
\usepackage{graphicx,color,epstopdf,subfigure}

\theoremstyle{plain}

\theoremstyle{remark}

\newcommand{\ux}{{\bf x}}
\newcommand{\un}{{\bf n}}
\newcommand{\uomega}{{\bf \Omega}}
\newcommand{\unabla}{{\bf \nabla}}

\begin{document}


\title[Non-classical transport and diffusion]
{The Non-Classical Boltzmann Equation, and DIffusion-Based Approximations to the Boltzmann Equation}

\author{Martin Frank}
\address[Martin Frank]{RWTH Aachen University,
Schinkelstrasse 2 \\ 52062 Aachen \\ Germany}
\email{frank@mathcces.rwth-aachen.de}

\author{Kai Krycki}
\address[Kai Krycki]{RWTH Aachen University,
Schinkelstrasse 2 \\ 52062 Aachen \\ Germany}
\email{krycki@mathcces.rwth-aachen.de}

\author{Edward W. Larsen}
\address[Edward W. Larsen]{Department of Nuclear Engineering and Radiological Sciences\\ University of Michigan\\ Ann Arbor, Michigan 48109, USA}
\email{edlarsen@umich.edu}

\author{Richard Vasques}
\address[Richard Vasques]{RWTH Aachen University,
Schinkelstrasse 2 \\ 52062 Aachen \\ Germany}
\curraddr{Universidade Federal do Rio Grande do Sul, PROMEC - School of Engineering, Osvaldo Aranha, 99 - 4o andar; 90046-900 Porto Alegre, RS, Brazil}
\email{richard.vasques@fulbrightmail.org}

\subjclass[2000]{}

\thanks{}

\begin{abstract}
We show that several diffusion-based approximations (classical diffusion or $SP_1$, $SP_2$, $SP_3$) to the linear Boltzmann equation can (for an infinite, homogeneous medium) be represented \emph{exactly} by a non-classical transport equation. 
As a consequence, we indicate a method to solve diffusion-based approximations to the Boltzmann equation via Monte Carlo, with only statistical errors - no truncation errors. 
\end{abstract}

\maketitle

\section{Introduction}
%
In the classical theory of linear particle transport, the total cross section $\Sigma_t$ is independent of the path-length $s$ (the distance traveled by the particle since its previous interaction), and of the direction of flight $\uomega$. In this case, the probability density function for a particle's distance-to-collision is given by an exponential:
\begin{equation}
p(s) = \Sigma_t e^{-\Sigma_t s}.
\end{equation}

However, in certain inhomogeneous random media in which the locations of the scattering centers are spatially correlated, the particle flux will experience a non-exponential attenuation law. A ``non-classical" theory for this type of transport problem was recently introduced \cite{larsen_07}, with the assumption that the positions of the scattering centers are
correlated but independent of direction. In the case of isotropic scattering, the non-classical linear Boltzmann equation is writen as
\begin{equation}
\label{eq:1}
\begin{split}
\frac{\partial\psi}{\partial s}(\ux,\uomega,s) &+ \uomega\cdot\unabla \psi(\ux,\uomega,s) + \Sigma_t(s)\psi(\ux,\uomega,s)\\
&= \frac{\delta(s)}{4\pi}\left[ c\int_{4\pi}\int_0^\infty \Sigma_t(s')\psi(\ux,\uomega',s')ds' d\Omega' + Q(\ux) \right].
\end{split}
\end{equation}
Here, $c$ is the scattering ratio (probability of scattering), and $Q(\bf x)$ is a source. The path length distribution
\begin{equation}\label{eq3}
p(s) = \Sigma_t(s)e^{-\int_0^s \Sigma_t(s')ds'}
\end{equation}
does not have to be exponential. If $p(s)$ is exponential, Eq.\ \eqref{eq:1} reduces to the classical Boltzmann equation for the classic angular flux 
\begin{equation}
\psi(\ux,\uomega) = \int_0^\infty \psi(\ux,\uomega,s)ds.
\end{equation} 
A full derivation of this non-classical linear Boltzmann equation
and its asymptotic diffusion limit can be found in \cite{larsen_11}, along with numerical results for an application in 2-D pebble bed reactor (PBR) cores. Existence and
uniqueness of solutions, as well as their convergence to the diffusion equation, are
rigorously discussed in \cite{frank_10}. The non-classical theory was extended in \cite{vasques_13a} to include angular-dependent path-length distributions, in order to investigate anisotropic diffusion of neutrons in 3-D PBR cores \cite{vasques_13b,vasques_13c}. Furthermore, a similar kinetic equation with path-length as an independent variable has been rigorously derived for the periodic Lorentz gas in a series of papers by Golse et al.\ (cf.\ \cite{golse_12} for a review) as well as Marklof \& Str\"ombergsson (cf.\  \cite{marklof_11}).

In this paper we do not (directly) deal with a random medium. Instead, we show that by selecting $\Sigma_t(s)$ properly, Eq.\ \eqref{eq:1} can be converted to an integral equation for the scalar flux
\begin{equation}
\label{eq:4}
\phi_0(\ux) = \int_{4\pi} \psi(\ux,\uomega)d\Omega,
\end{equation}
which is \emph{identical} to the integral equation that can be constructed for several diffusion-based approximations to the \emph{classic} Boltzmann transport equation
\begin{equation}
\label{eq:5}
\uomega\cdot\unabla \psi(\ux,\uomega) + \Sigma_t \psi(\ux,\uomega) = \frac{\Sigma_s}{4\pi}\int_{4\pi}\psi(\ux,\uomega)d\Omega'+ \frac{Q(\ux)}{4\pi}.
\end{equation}

In other words, we show that for an infinite homogeneous medium
in which
\begin{itemize}
\item[(i)] $\Sigma_s<\Sigma_t$,
\item[(ii)] $Q(\ux)\to 0$ as $|\ux|\to\infty$,
\item[(iii)] $\psi(\ux,\uomega)\to 0$ as $|\ux|\to\infty$,
\end{itemize}
the classical linear Boltzmann equation (Eq.\ \eqref{eq:5}) and several of its diffusion-based approximations can \emph{all} be exactly represented by the non-classical Boltzmann equation \eqref{eq:1} with a correctly chosen $\Sigma_t(s)$. Moreover, the exact definition of $\Sigma_t(s)$ for each method can be determined (semi-)analytically.

To describe the diffusion-based approximations to the transport Eq.\ \eqref{eq:5}, we
integrate Eq.\ \eqref{eq:5} over $\uomega$, defining $\phi(\ux)$ by Eq.\ \eqref{eq:4}, and 
$$
\phi_1(\ux) = \int\uomega\psi(\ux,\uomega)d\Omega = \text{current},
$$
we obtain the exact balance equation
\begin{equation}
\uomega\cdot\unabla\phi_1(\ux) + \Sigma_t \phi_0(\ux) = \Sigma_s \phi_0(\ux) + Q(\ux).
\end{equation}

Diffusion-based methods invoke a closure relation, which expresses $\phi_1$ in terms of $\phi_0$. The classic diffusion approximation invokes \emph{Fick's Law}:
\begin{equation}
\phi_1(\ux) = -\frac{1}{3\Sigma_t}\unabla\phi_0(\ux)
\end{equation}
to give (with $\Sigma_t -\Sigma_s = \Sigma_a$):
\begin{equation}
\label{eq:9}
-\frac{1}{3\Sigma_t}\nabla^2 \phi_0(\ux) + \Sigma_a \phi_0(\ux) = Q(\ux).
\end{equation}

The classic diffusion equation has been generalized to the hierarchy of $SP_N$ equations. A recent and complete review on these equations is \cite{mcclarren_11}. The $SP_N$ equations were first derived by Gelbard \cite{Gel60,Gel61,Gel62} in an ad-hoc way. Theoretical justifications were presented later \cite{LarMorMcG93, TomLar96}.

In the $SP_2$ approximation, Eq.\ \eqref{eq:9} is generalized to:
\begin{equation}
-\frac{1}{3\Sigma_t}\nabla^2\left[ \phi_0 + \frac{4}{5\Sigma_t}(\Sigma_a\phi_0-Q) \right] + \Sigma_a\phi_0 = Q.
\end{equation}

In the $SP_3$ approximation, Eq.\ \eqref{eq:9} is generalized to the system:
\begin{subequations}
\begin{align}
-\frac{1}{3\Sigma_t}\nabla^2(\phi_0+2\phi_2)+\Sigma_a\phi_0 &= Q,\\
-\frac{9}{35\Sigma_t}\nabla^2\phi_2+\Sigma_t\phi_2 &= \frac{2}{5}(\Sigma_a\phi_0-Q).
\end{align}
\end{subequations}


The work in this paper accomplishes the following:
\begin{itemize}
\item[(i)] It demonstrates (for an infinite homogeneous medium) that the original Boltzmann equation \emph{and} the above-mentioned diffusion-based approximations to this equation are all special cases of the non-classical Boltzmann equation. This sheds some new light on the various diffusion approximations.
\item[(ii)] Since the non-classical Boltzmann equation \eqref{eq:1} can be solved by Monte Carlo methods, the results in these notes show how to solve diffusion-based approximations to the Boltzmann equation via Monte Carlo, with only statistical errors - no truncation errors. To our knowledge, this has not been done before.
\item[(iii)] In connection with (ii), if $p(s)ds =$ the probability that a particle will experience a collision between path length $s$ and $s+ds$ (since the previous collision), then the distance to collision $s$ can be sampled by inverse transform sampling from the cumulative distribution function
\begin{equation}
\label{eq:6}
\xi = \int_0^s p(s') ds'.
\end{equation}
In this paper, we show that for all the diffusion-based methods considered, the forms of $p(s)$ are such that Eq.\ \eqref{eq:6} can be \emph{explicitly solved} for $s$ in terms of $\xi =$ (computer-generated) random number, uniformly distributed between 0 and 1. This makes the possibility of using Monte Carlo to solve these equations much more realistic.
\item[(iv)] Finally, this work shows that non-classical transport processes have been widely used for many years, without explicit awareness of this. It may make it possible to consider other unknown-at-present applications of non-classical transport for problems in which the assumptions of classical transport are too limiting.
\end{itemize}

The remainder of this paper is organized as follows. In Section \ref{sec:2} we use the previous work on the non-classical Boltzmann equation to convert Eq.\ \eqref{eq:1} to an integral equation for the scalar flux $\phi_0$. In Section \ref{sec:3}, we use the Green's function for the diffusion operator
$$
-\nabla^2\phi+\Sigma_t^2\lambda^2\phi
$$
to convert Eq.\ \eqref{eq:9} into an integral equation for $\phi_0(\ux)$. By choosing $\Sigma_t(s)$ correctly, the integral equation obtained in Section \ref{sec:2} becomes identical to this (diffusion) integral equation. 
In Sections \ref{sec:5} and \ref{sec:6} we show that the SP$_2$ and SP$_3$ approximations to the classic Boltzmann equation can, like the standard diffusion approximation treated in Section \ref{sec:3}, be represented as non-classical transport equations. We conclude with a discussion in Section \ref{sec:7}.

\section{Integral Equation Formulation}\label{sec:2}
%

To simplify the notation, we define the scattering-plus-inhomogeneous source (the right-hand side of Eq.\ \eqref{eq:1}) by
\begin{subequations}
\begin{align}\label{eq14a}
S(\ux) &= c\int_{4\pi}\int_0^\infty \Sigma_t(s')\psi(\ux,\uomega',s')ds'd\Omega'+Q(\ux) \\
&= c\int_0^\infty \Sigma_t(s')\phi_0(\ux,s')ds' + Q(\ux) \nonumber \\
&= cf(\ux) + Q(\ux), \nonumber
\end{align}
where 
\begin{align}
\phi_0(\ux,s) &= \int_{4\pi} \psi(\ux,\uomega,s)d\Omega = \text{ non-classical scalar flux},\\
f(\ux) &= \int_0^\infty \Sigma_t(s')\phi_0(\ux,s')ds' = \text{ collision-rate density.}\label{eq:15}
\end{align}
\end{subequations}
Then Eq.\ \eqref{eq:1} can be written as
\begin{equation}
\frac{\partial\psi}{\partial s}(\ux,\uomega,s) + \uomega\cdot\unabla\psi(\ux,\uomega,s) + \Sigma_t(s)\psi(\ux,\uomega,s) = \frac{\delta(s)}{4\pi}S(\ux);
\end{equation}
or, equivalently, as
\begin{subequations}
\label{eq:17}
\begin{equation}
\frac{\partial\psi}{\partial s}(\ux,\uomega,s) + \uomega\cdot\unabla\psi(\ux,\uomega,s) + \Sigma_t(s)\psi(\ux,\uomega,s) = 0,
\end{equation}
and
\begin{equation}
\psi(\ux,\uomega,0) = \frac{S(\ux)}{4\pi}.
\end{equation}
\end{subequations}

Following \cite{larsen_11}, we use the method of characteristics to calculate the solution of Eqs.\ \eqref{eq:17}:
\begin{equation}
\psi(\ux,\uomega,s) = \frac{S(\ux-s\uomega)}{4\pi}e^{-\int_0^s \Sigma_t(s')ds'}.
\end{equation}
Operating on this equation by
$$
\int_0^\infty \Sigma_t(s)(\ \cdot\ )ds
$$
and using Eq.\ \eqref{eq3}, we obtain:
$$
\int_0^\infty \Sigma_t(s)\psi(\ux,\uomega,s)ds = \frac{1}{4\pi}\int_0^\infty S(\ux-s\uomega)p(s)ds.
$$
Now, operating by $\int_{4\pi}(\ \cdot\ )d\Omega$ and using Eq.\ (\ref{eq:15}), we get
$$
f(\ux) = \int_0^\infty \Sigma_t(s)\phi_0(\ux,s)ds = \frac{1}{4\pi}\int_0^\infty\int_{4\pi} S(\ux-s\uomega)p(s)d\Omega ds.
$$
Finally, we make the change of spatial variables from the 3-D spherical $(\uomega,s)$ to the 3-D Cartesian $\ux'$ defined by
\begin{subequations}
\begin{align}
\ux'&= \ux-s\uomega.
\end{align}
Then
\begin{align}
s &= |\ux'-\ux| = \text{ radial variable},\\
dV'&= dx'dy'dz' = s^2dsd\Omega,\\
dsd\Omega &= \frac{dV'}{s^2} = \frac{dV'}{|\ux'-\ux|^2},
\end{align}
\end{subequations}
and we obtain:
\begin{equation}
\label{eq:20a}
f(\ux) = \int\int\int S(\ux') \frac{p(|\ux'-\ux|)}{4\pi |\ux'-\ux|^2} dV',
\end{equation}
where $p(|\ux'-\ux|)$ and $S(\ux)$ are given by Eqs.\ \eqref{eq3} and \eqref{eq14a}, respectively.

Finally, for classic particle transport (in which $\Sigma_t$ is independent of $s$), we have
\begin{subequations}\label{eq:21}
\begin{align}
\label{eq:21c}
f(\ux) &= \int\int\int S(\ux') \frac{\Sigma_t e^{-\Sigma_t |\ux-\ux'|}}{4\pi|\ux-\ux'|^2}dV' \\
&= \int\int\int [cf(\ux')+Q(\ux')] \frac{\Sigma_t e^{-\Sigma_t |\ux-\ux'|}}{4\pi|\ux-\ux'|^2}dV',\nonumber
\end{align}
which is equivalent to the classic integral equation for the scalar flux. Hence, Eq.\ \eqref{eq:6} for sampling $s$ yields
\begin{align}
\xi = \int_0^s p(s') ds' = \int_0^s \Sigma_t e^{-\Sigma_t s'}ds' = 1-e^{-\Sigma_t s},
\end{align}
which we can rewrite as
\begin{equation}
s = -\frac{1}{\Sigma_t}\ln(1-\xi).
\end{equation}
\end{subequations}
Eqs.\ \eqref{eq:21} are all standard results, which demonstrate the fairly obvious fact that when $\Sigma_t(s) = \Sigma_t = \text{constant}$, the non-classical Boltzmann equation reduces to the standard Boltzmann equation.
Next, we derive similar results for diffusion-based approximations to Eq.\ \eqref{eq:5}. These results are \emph{not} standard.

\section{Classic Diffusion}\label{sec:3}
\begin{figure}[h]
\centering\includegraphics[width=0.8\linewidth]{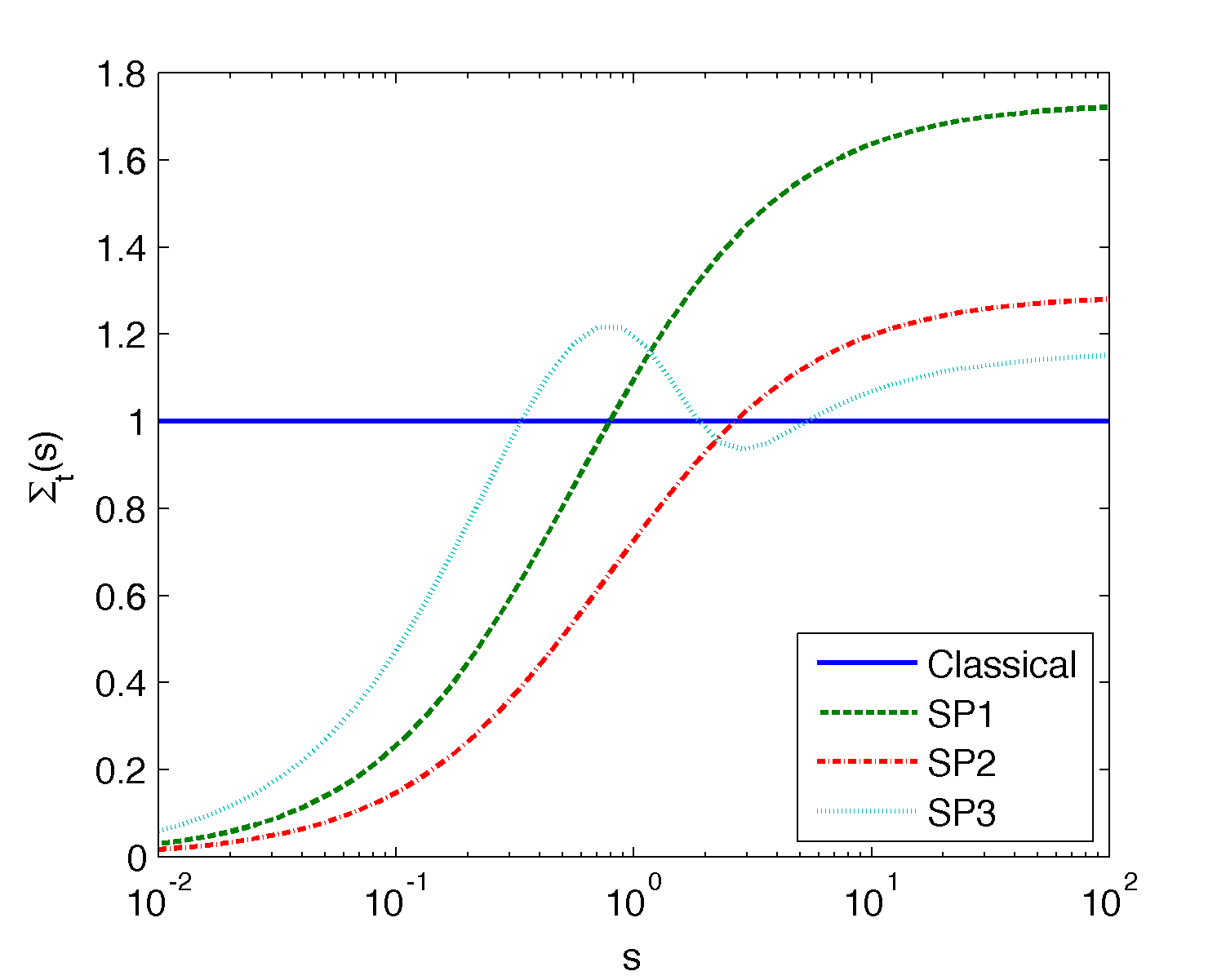}
\caption{Cross-section $\Sigma_t$ as a function of path length $s$. Comparison of classical transport and diffusion approximations.}
\label{fig:Sigmat}
\end{figure}

\begin{figure}[h]
\centering\includegraphics[width=0.8\linewidth]{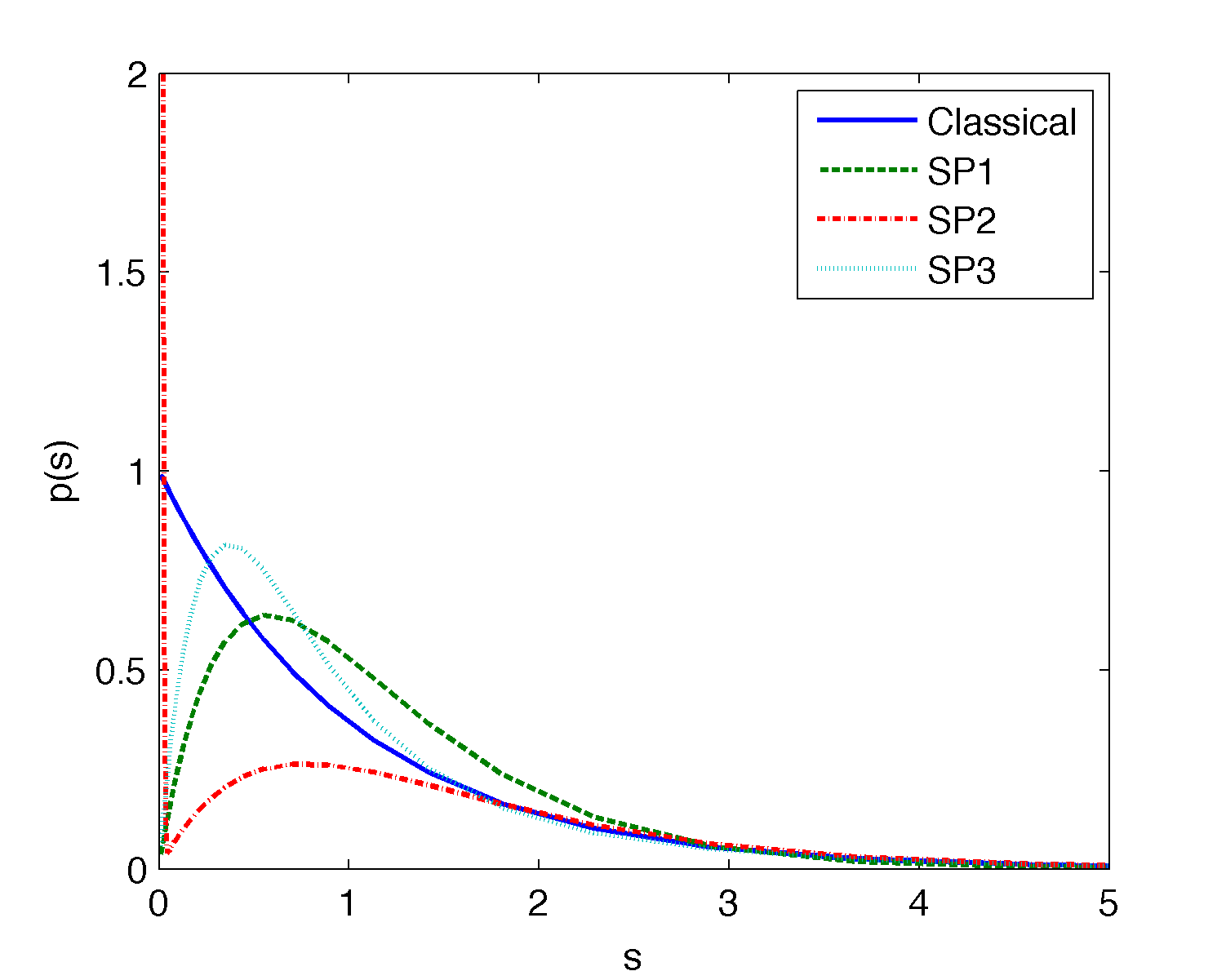}
\caption{Path-length probability density function $p(s)$. Comparison of classical transport and diffusion approximations.}
\label{fig:p}
\end{figure}

%
Let us repeat the statements in the introduction on how diffusion is typically derived:
Integrating Eq.\ \eqref{eq:5} over $\uomega$, defining $\phi(\ux)$ by Eq.\ \eqref{eq:4}, and 
$$
{\bf\phi_1}(\ux) = \int\uomega\psi(\ux,\uomega)d\Omega = \text{current},
$$
we obtain the exact balance equation
\begin{equation}
\unabla\cdot{\bf\phi_1}(\ux) + \Sigma_t \phi(\ux) = \Sigma_s \phi(\ux) + Q(\ux).
\end{equation}
Diffusion-based methods invoke a closure relation, which expresses ${\bf\phi_1}$ in terms of $\phi$. The classic diffusion aproximation invokes \emph{Fick's Law}:
\begin{equation}
{\bf\phi_1}(\ux) = -\frac{1}{3\Sigma_t}\unabla\phi_0(\ux)
\end{equation}
to give
\begin{equation}
\label{eq:9a}
-\frac{1}{3\Sigma_t}\nabla^2\phi_0(\ux) + \Sigma_t\phi_0(\ux) = \Sigma_s\phi_0(\ux) + Q(\ux),
\end{equation}
which is the classic diffusion approximation to Eq.\ \eqref{eq:5}. (Equation \eqref{eq:9a} is commonly written in the form
\begin{equation}
\label{eq:22}
-\frac{1}{3\Sigma_t}\nabla^2 \phi_0(\ux) + \Sigma_a \phi_0(\ux) = Q(\ux),
\end{equation}
with $\Sigma_a=\Sigma_t -\Sigma_s$.) If we define $S(\ux) = \Sigma_s\phi_0(\ux)+Q(\ux)$, we can rewrite 
Eq.\ \eqref{eq:9a} as:
\begin{equation}
\label{eq:23}
-\nabla^2\phi_0(\ux) + \Sigma_t^2\lambda^2\phi_0(\ux) = 3\Sigma_t S(\ux),
\end{equation}
where 
$$
\lambda^2 = 3.
$$
The Green's function for the operator on the left hand side of Eq.\ \eqref{eq:23} is:
\begin{equation}
\label{eq:24}
G(|\ux-\ux'|) = \frac{e^{-\sqrt{3}\Sigma_t |\ux-\ux'|}}{4\pi |\ux-\ux'|}.
\end{equation}
Therefore, we can manipulate Eq.\ \eqref{eq:23} for $\phi_0(\ux)$ by taking
\begin{align*}
\phi_0(\ux) &= \int\int\int G(|\ux-\ux'|) 3\Sigma_t S(\ux') dV' \\
&= \int\int\int \frac{3\Sigma_te^{-\sqrt{3}\Sigma_t |\ux-\ux'|}}{4\pi |\ux-\ux'|} S(\ux') dV' \\
&= \int\int\int \frac{3\Sigma_t |\ux-\ux'| e^{-\sqrt{3}\Sigma_t |\ux-\ux'|}}{4\pi |\ux-\ux'|^2} S(\ux') dV'.
\end{align*}
Now we multiply by $\Sigma_t$ to obtain the collision rate density $f=\Sigma_t \phi_0$:
\begin{equation*}
\Sigma_t \phi_0 = \int\int\int \frac{3\Sigma_t^2 |\ux-\ux'| e^{-\sqrt{3}\Sigma_t |\ux-\ux'|}}{4\pi |\ux-\ux'|^2} S(\ux')dV'.
\end{equation*}
This result agrees with Eq.\ \eqref{eq:20a} iff
\begin{equation}
\label{eq:25}
p(s) = 3\Sigma_t^2 s e^{-\sqrt{3} \Sigma_t s}.
\end{equation}
It is easily confirmed that
$$
\int_0^\infty 3\Sigma_t^2 s e^{-\sqrt{3} \Sigma_t s} ds = \int_0 ^\infty \sqrt{3}\Sigma_ts e^{-\sqrt{3}\Sigma_t s} d(\sqrt{3}\Sigma_t s) = 1,
$$
so Eq.\ \eqref{eq:25} does describe a distribution function. $\Sigma_t(s)$ is given by
\begin{equation}
\label{eq:26}
\Sigma_t(s) = \frac{p(s)}{\int_s^\infty p(s')ds'} = \frac{3\Sigma_t^2s}{1+\sqrt{3}\Sigma_t s} = \sqrt{3}\Sigma_t\frac{\sqrt{3}\Sigma_t s}{1+\sqrt{3}\Sigma_t s}.
\end{equation}
Therefore: the non-classical transport equation reproduces the classic diffusion approximation Eq.\ \eqref{eq:22} to Eq.\ \eqref{eq:1} if $\Sigma_t(s)$ and $p(s)$ are defined by Eqs.\ \eqref{eq:25} and \eqref{eq:26}. Both functions are shown in Figure \ref{fig:Sigmat} and Figure \ref{fig:p}, respectively, at the end of the paper.

\begin{figure}[h]
\centering\includegraphics[width=0.8\linewidth]{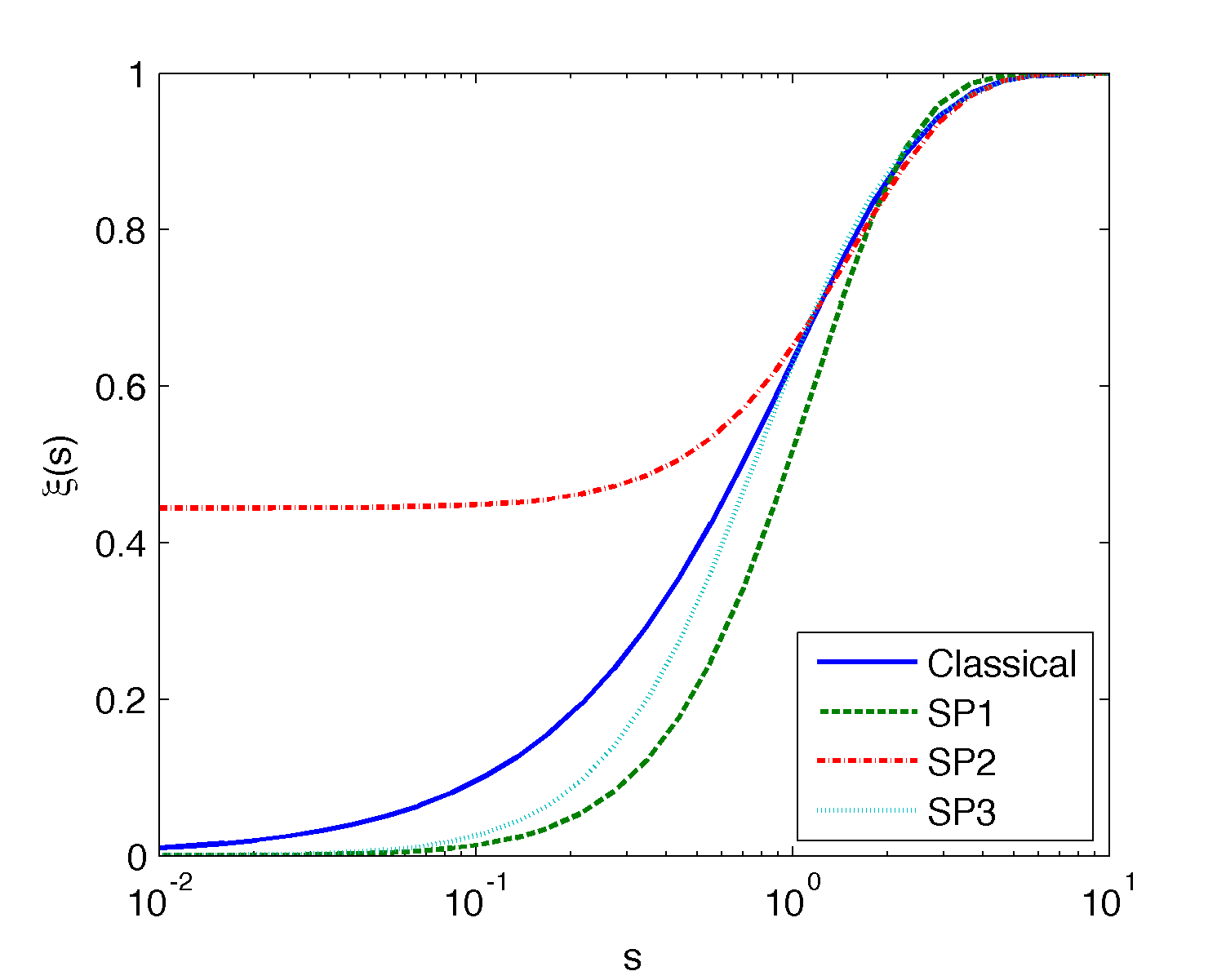}
\caption{Path-length cumulative distribution function $\xi(s)$. Comparison of classical transport and diffusion approximations.}
\label{fig:xi}
\end{figure}

We note that the mean distance to collision (the mean free path) is:
\begin{equation}
\label{eq:28}
\bar s = \int_0^\infty sp(s) ds = \int_0^\infty s 3\Sigma_t^2 s e^{-\sqrt{3}\Sigma_t s} ds = \frac{2}{\sqrt{3}\Sigma_t}.
\end{equation}
This, of course, is greater than $\bar s = \Sigma_t^{-1}$ for the original transport equation. Also, $s$ can be sampled by
$$
\xi = \int_0^s p(s') ds' = \int_0^s 3\Sigma_t^2s'e^{-\sqrt{3}\Sigma_t s'} ds' = 1-(1+\sqrt{3}\Sigma_t s)e^{-\sqrt{3}\Sigma_t s}
$$
and thus
\begin{subequations}
\begin{equation}
s = \frac{1}{\sqrt{3}\Sigma_t}f^{-1} (\xi),
\end{equation}
where
\begin{equation}
\label{eq:29b}
f(z) = (1+z)e^{-z}.
\end{equation}
\end{subequations}
The function $f(z)$ is monotonic decreasing for $0<z<\infty$, taking values in $0<f(z)<1$. The inverse $f^{-1}(\xi)$ can be precomputed by table, or interpolated, or otherwise computed; we will not consider this here. We show $\xi(s)$ in Figure \ref{fig:xi} at the end of the paper.

\section{Simplified $P_2$ ($SP_2$)}\label{sec:5}
%
The SP$_2$ approximation to Eq.\ \eqref{eq:5} is:
\begin{equation}
- \frac{1}{3\Sigma_t} \nabla^2\left[\phi_0 + \frac{4}{5\Sigma_t}(\Sigma_a\phi_0 - Q)\right] + \Sigma_a\phi_0 = Q .
\end{equation}
Equivalently, 
\begin{equation*}
\begin{split}
- \frac{1}{3\Sigma_t} \nabla^2\left(1+\frac{4\Sigma_a}{5\Sigma_t}\right)\phi_0 + \Sigma_t  \phi_0
&=(\Sigma_s \phi_0 + Q) - \frac{4}{15\Sigma_t^2} \nabla^2 Q \\
 &= S - \frac{4}{15\Sigma_t^2} \nabla^2 [(\Sigma_s \phi_0 + Q) - \Sigma_s \phi_0]\\
&= S - \frac{4}{15\Sigma_t^2} \nabla^2 S + \frac{4\Sigma_s}{15\Sigma_t^2} \nabla^2 \phi_0 ,
\end{split}
\end{equation*}
where $S = (\Sigma_s\phi_0+Q)$.
Bringing the $\nabla^2 \phi_0$ term to the left side, we obtain:
$$
- \nabla^2 \left(\frac{1}{3\Sigma_t} + \frac{4\Sigma_a}{15\Sigma_t^2} + \frac{4\Sigma_s}{15\Sigma_t^2}\right) \phi_0 + \Sigma_t \phi_0 = S -\frac{4}{15\Sigma_t^2} \nabla^2 S ,
$$
or, since
$$
\frac{1}{3\Sigma_t} + \frac{4}{15} \frac{\Sigma_a + \Sigma_s}{\Sigma_t^2} = 
\frac{1}{3\Sigma_t} + \frac{4}{15 \Sigma_t} = \frac{3}{5 \Sigma_t} , 
$$
we have
\begin{equation}
\label{eq:42}
- \frac{3}{5 \Sigma_t} \nabla^2 \phi_0 + \Sigma_t\phi_0 = S - \frac{4}{15\Sigma_t^2} \nabla^2 S. 
\end{equation}
Multiplying by $\frac{5\Sigma_t}{3}$, we obtain
\begin{equation*}
\begin{split}
- \nabla^2\phi_0 + \left(\frac{5}{3} \Sigma_t^2\right) \phi_0 
&= \frac{5\Sigma_t}{3} S - \frac{4}{9\Sigma_t}\nabla^2 S\\
&=  
\frac{5\Sigma_t}{3} S + \frac{4}{9\Sigma_t} \left(-\nabla^2 S + \frac{5}{3} \Sigma_t^2 S- \frac{5}{3} \Sigma_t^2 S^2\right) .
\end{split}
\end{equation*}
Defining
$$
\lambda^2 = \frac{5}{3} ,
$$
we obtain:
\begin{equation}
\begin{split}
(-\nabla^2 + \Sigma_t^2\lambda^2) \phi_0 
&=\left(\frac{5\Sigma_t}{3} - \frac{4}{9\Sigma_t}\frac{5}{3}\Sigma_t^2\right)S + \frac{4}{9\Sigma_t} (-\nabla^2 + \Sigma_t^2\lambda^2)S\\
&= \frac{25}{27}\Sigma_t S + \frac{4}{9\Sigma_t}(-\nabla^2+\Sigma_t^2\lambda^2) S.
\end{split}
\end{equation}
Using the Green's Function [Eq.\ \eqref{eq:24}] for $ -\nabla^2 + \Sigma_t^2 \lambda^2 $, we get:
 $$
\phi_0 (\ux)= \frac{25}{27}\Sigma_t \int\int\int G S dV' + \frac{4}{9\Sigma_t} S ,
$$
or,
\begin{equation}
\Sigma_t\phi_0 =\frac{5}{9} {(\Sigma_t\lambda)}^2 \int\int\int G SdV' +\frac{4}{9}S ,
\end{equation}
where
\begin{equation}
G(s) = \frac{e^{-\Sigma_t\lambda s}}{4\pi s} . 
\end{equation}

Now, we use the identity
\begin{equation*}
\begin{split}
S(\ux) = \int_0^\infty S(\ux + s \uomega) \delta(s)ds
&= \frac{1}{4\pi} \int_{4\pi}\int\delta(s) S(\ux +s\uomega)dsd\Omega\\
&= \int_{4\pi}\int\frac{\delta(|\ux - \ux'|)}{4\pi} \frac{S(\ux')}{|\ux - \ux'|^2} dV' ,
\end{split}
\end{equation*}
[where $\ux'=\ux+s\uomega$, $|\ux-\ux'|=s$, $s^2dsd\Omega=dV'$] to obtain
from Eq.\ \eqref{eq:42}:
\begin{equation}
\begin{split}
\Sigma_t \phi_0 (\ux) = \frac{5}{9} \int\int\int &\frac{\Sigma_t^2 \lambda^2 |\ux-\ux'| e^{-\Sigma_t \lambda |\ux - \ux'|}}{4 \pi |\ux-\ux'|^2} S(\ux')dV' \\
&+ \frac{4}{9} \int\int\int \frac{\delta(|\ux-\ux'|)}{4\pi|\ux-\ux'|^2} S(\ux')dV' .
\end{split}
\end{equation}
This implies that for the SP$_2$ equation,
\begin{subequations}\label{eq:47}
\begin{equation}
p(s)=\frac{5}{9} \Sigma_t^2 \lambda^2s e^{-\Sigma_t\lambda s} + \frac{4}{9}\delta(s),
\end{equation}
where
\begin{equation}
\lambda=\sqrt{\frac{5}{3}}.
\end{equation}
\end{subequations}

Thus, with probability $\frac{4}{9}$, a particle that scatters at a point $\ux$ undergoes its next ``collision" at the same point. Each time a particle experiences a collision (even if it has not moved), it has the probability of being absorbed (with probability $\Sigma_a/\Sigma_t$).

From Eq.\ \eqref{eq:26}, we have for $s>0$
\begin{equation}
\label{eq:48}
\Sigma_t(s) = \frac{p(s)}{\int_s^\infty p(s')ds'} = \frac{\Lambda^2 s e^{-\Lambda s}}{\int_s^\infty \Lambda^2 s' e^{-\Lambda s'} ds'} = \frac{\Lambda^2 s}{1+\Lambda s},
\end{equation}
where $\Lambda = \Sigma_t\lambda$.
Also, for $s\approx 0$, Eqs.\ \eqref{eq:25} and \eqref{eq:47} give
\begin{equation}
\label{eq:49}
\Sigma_t(s)\approx\frac{4}{9}\delta(s).
\end{equation}
This agrees with the physical interpretation that
$$
\Sigma_t(0)ds = \frac{4}{9} = \text{the probability that a particle at $s=0$ will experience a collision.}
$$
Equations \eqref{eq:48} and \eqref{eq:49} can be written more compactly as
\begin{equation}
\label{eq:50}
\Sigma_t(s) = \frac{\frac{4}{9}\delta(s)+\Lambda^2s}{1+\Lambda s}.
\end{equation}

The mean free path is:
\begin{equation}
\label{eq:51}
\begin{split}
\bar s &= \int_0^\infty sp(s)ds = \int_0^\infty s\left[ \frac{5}{9}\Lambda^2 s e^{-\Lambda s}+\frac{4}{9}\delta(s)\right]ds = \frac{10}{9\Lambda} \\
&= \frac{10}{9}\sqrt{\frac{3}{5}}\frac{1}{\Sigma_t} = \sqrt{\frac{20}{27}}\frac{1}{\Sigma_t}.
\end{split}
\end{equation}
This result is less than the physically correct $\frac{1}{\Sigma_t}$.

Finally, the distance-to-collision can be sampled using
\begin{subequations}\label{eq:52}
\begin{equation}
\label{eq:52a}
\begin{split}
\xi &= \int_0^s p(s')ds' = \int_0^s\left[\frac{4}{9}\delta(s')+\frac{5}{9}\Lambda^2s'e^{-\Lambda s'}\right]ds'\\
& = \frac{4}{9} + \frac{5}{9}\left[1-(1+\Lambda s)e^{-\Lambda s}\right] = 1-\frac{5}{9}f(\Lambda s),
\end{split}
\end{equation}
where
\begin{equation}
\label{eq:52b}
f(z) = (1+z)e^{-z}
\end{equation}
\end{subequations}
was introduced earlier [in Eq.\ \eqref{eq:29b}]. 
Thus, for $0\leq\xi\leq\frac{4}{9}$, $s=0$. For $\xi>\frac{4}{9}$, Eq.\ \eqref{eq:52a} gives
$$
\frac{5}{9}f(\Lambda s) = 1-\xi 
$$
$$
\Rightarrow \xi = \sqrt{\frac{3}{5}}\frac{1}{\Sigma_t} f^{-1}\left(\frac{9}{5}(1-\xi)\right).
$$
Equivalently,
\begin{equation}
\label{eq:53}
s = \left\{
\begin{tabular}{cl}
$0$ & $0\leq \xi\leq \frac{4}{9}$,\\
$\sqrt{\frac{3}{5}}\frac{1}{\Sigma_t} f^{-1}\left(\frac{9}{5}(1-\xi)\right)$ & $\frac{4}{9}<\xi\leq1$.
\end{tabular}
\right.
\end{equation}

Again, Eq.\ \eqref{eq:53} states that with probability $\frac{4}{9}$, a particle will suffer its next collision at the precise location of the previous one. Each time a particle experiences a collision (whether it moves or not) it is subject to absorption, with probability $\Sigma_a/\Sigma_t$. The functions $\Sigma_t(s)$, $p(s)$, and $\xi(s)$ are shown in Figures \ref{fig:Sigmat}-\ref{fig:xi}.

\section{Simplified $P_3$ ($SP_3$)}\label{sec:6}
%
The SP$_3$ approximation to Eq.\ \eqref{eq:5} consists of the following two coupled equations:
\begin{subequations}\label{eq:54}
\begin{align}
-\frac{1}{3\Sigma_t}\nabla^2(\phi_0+2\phi_2)+\Sigma_t\phi_0 = S,\\
-\frac{9}{35\Sigma_t}\nabla^2\phi_2+\Sigma_t\phi_2 = \frac{2}{5}(\Sigma_t\phi_0-S),
\end{align}
where
\begin{align}
S(\ux) = \Sigma_s\phi_0(\ux)+Q(\ux).
\end{align}
\end{subequations}
To demonstrate that the SP$_3$ equations can be represented as a non-classical transport process, we must calculate the Green's function for Eqs.\ \eqref{eq:54}. These functions satisfy:
\begin{subequations}\label{eq:55}
\begin{align}
-\frac{1}{3\Sigma_t}\nabla^2(G_0+2G_2)+\Sigma_tG_0 = \delta(\ux)\\
-\frac{9}{35\Sigma_t}\nabla^2G_2+\Sigma_tG_2-\frac{2}{5}\Sigma_tG_0 = -\frac{2}{5}\delta(\ux),
\end{align}
\end{subequations}
where $G_0$ and $G_2$ are functions of $r=|\ux|$, and when found, they enable Eqs.\ \eqref{eq:54} to be written:
\begin{subequations}
\begin{align}
\phi_0(\ux) = \int\int\int G_0(|\ux-\ux'|)S(\ux')dV',\label{eq:56a}\\
\phi_2(\ux) = \int\int\int G_2(|\ux-\ux'|)S(\ux')dV'\label{eq:56b}.
\end{align}
\end{subequations}
Equation \eqref{eq:56b} is not needed, but Eq.\ \eqref{eq:56a} is needed to show the desired result.

We know that
\begin{subequations}
\begin{align}
\label{eq:57a}
G(r) = \frac{e^{-\Sigma_t\lambda r}}{4\pi r}
\end{align}
is the Green's function for the operator $-\nabla^2+\Sigma_t^2\lambda^2$, i.e. it satisfies:
\begin{align}
\label{eq:57b}
-\nabla^2G+\Sigma_t^2\lambda^2G = \delta(\ux).
\end{align}
\end{subequations}
More specifically,
\begin{itemize}
\item[1.] For $r>0$, $G(r)$ satisfies
\begin{subequations}\label{eq:58}
\begin{align}
-\frac{1}{r^2}\frac{\partial}{\partial r} r^2\frac{\partial G}{\partial r} + \Sigma_t^2\lambda^2G = 0.
\end{align}
\item[2.] Also, if we integrate Eq.\ \eqref{eq:57b} over $|x|\leq\varepsilon$,
\begin{align}\label{eq:58b}
-\int_{|\ux|\leq\varepsilon}\unabla\cdot\unabla GdV = -\int_{|\ux|=\varepsilon}{\un}\cdot\unabla GdS = -(4\pi\varepsilon^2)\frac{\partial G}{\partial r}(\varepsilon),
\end{align}
and let $\varepsilon \rightarrow 0$, we get
\begin{align}
\lim_{\varepsilon\rightarrow 0}\left(-4\pi\varepsilon^2\frac{\partial G}{\partial r}(\varepsilon)\right) = 1.
\end{align}
\end{subequations} 
The right side of Eq.\ \eqref{eq:58b} is the rate at which the $\delta$-function source emits particles at $\ux = 0$. The left side of this equation is the net rate at which particles leak away from the point $\ux =0$.
\end{itemize}
It is easily verified that $G(r)$, defined by Eq.\ \eqref{eq:57a} satisfies both Eqs.\ \eqref{eq:58}.

To solve Eqs.\ \eqref{eq:55}, we seek two functions, $G_0(r)$ and $G_2(r)$, satisfying:
\begin{itemize}
\item[1.] For $0<r<\infty$,
\begin{subequations}\label{eq:59}
\begin{align}
-\frac{1}{3\Sigma_t}\frac{1}{r^2}\frac{\partial}{\partial r}r^2\frac{\partial}{\partial r}\left(G_0+2G_2\right)+\Sigma_tG_0 = 0,\\
-\frac{9}{35\Sigma_t}\frac{1}{r^2}\frac{\partial}{\partial r}r^2\frac{\partial}{\partial r}G_2 + \Sigma_tG_2-\frac{2}{5}\Sigma_tG_0 = 0,
\end{align}
\end{subequations}
\item[2.]
\begin{subequations}\label{eq:60}
\begin{align}
-\frac{1}{3\Sigma_t}\lim_{\varepsilon\rightarrow 0}\left[(4\pi\varepsilon^2)
\left(\frac{\partial G_0}{\partial r}(\varepsilon)+2\frac{\partial G_2}{\partial r}(\varepsilon)\right)
\right] = 1,\\
-\frac{9}{35\Sigma_t}\lim_{\varepsilon\rightarrow 0}\left[(4\pi\varepsilon^2)\frac{\partial G_2}{\partial r}(\varepsilon)\right] = -\frac{2}{5}.
\end{align}
\end{subequations}
\end{itemize}
Note that Eqs.\ \eqref{eq:60} were obtained by operating on Eqs.\ \eqref{eq:55} by $\displaystyle{\lim_{\varepsilon\rightarrow 0}}\int_{|\ux|\leq\varepsilon}(\cdot)dV.$

To satisfy Eqs.\ \eqref{eq:59}, we seek solutions of these equations of the form:
\begin{subequations}\label{eq:61}
\begin{align}
G_0(r) = \frac{e^{-\Sigma_t\lambda r}}{4\pi r},\\
G_2(r) = a\frac{e^{-\Sigma_t\lambda r}}{4\pi r},
\end{align}
\end{subequations}
where $\lambda$ and $a$ are constants to be determined. Using
$$
\nabla^2\left(\frac{e^{-\Sigma_t \lambda r}}{4\pi r}\right) = \Sigma_t^2\lambda^2\left(\frac{e^{-\Sigma_t\lambda r}}{4\pi r}\right),
$$
equations \eqref{eq:59} yield:
$$
-\frac{1}{3\Sigma_t}[\Sigma_t^2\lambda^2+2a\Sigma_t^2\lambda^2]+\Sigma_t = 0,
$$
$$
-\frac{9}{35\Sigma_t}[\Sigma_t^2\lambda^2 a]+\Sigma_t a-\frac{2}{5}\Sigma_t = 0,
$$
or:
$$
-\frac{1}{3}(1+2a)\lambda^2 + 1=0,
$$
$$
-\frac{9}{35}a\lambda^2+a = \frac{2}{5}.
$$
The second of these equations gives
\begin{subequations}
\begin{align}
a=\frac{14}{35-9\lambda^2},\label{eq:62a}
\end{align}
and then the first gives:
\begin{align}
\label{eq:62b}
-\frac{1}{3}\left(1+\frac{28}{35-9\lambda^2}\right)\lambda^2+1=0.
\end{align}
\end{subequations}
Simple manipulations of Eq.\ \eqref{eq:62b} give:
$$
0=3\lambda^4-30\lambda^2+35.
$$
This equation has two solutions:
$$
(\lambda^{\pm})^2 = 5\pm 2\sqrt{\frac{10}{3}}\approx 5\pm 3.651484.
$$
Thus,
\begin{subequations}
\begin{align}
(\lambda^+)^2 = 8.651482,\\
(\lambda^-)^2 = 1.348516.
\end{align}
\end{subequations}
Taking the square roots, we get
\begin{subequations}\label{eq:64}
\begin{align}
\lambda^+ = 2.941340,\label{eq:64a}\\
\lambda^- = 1.161256.\label{eq:64b}
\end{align}
\end{subequations}
Note: For the classic diffusion equation treated in Section \ref{sec:3}, $\lambda=\sqrt{3}$, and
$$
\mu = \frac{1}{\lambda}=\frac{1}{\sqrt{3}} = 0.577350
$$
is the positive value of $\mu$ in the S$_2$ Gauss-Legendre quadrature set. For Eqs.\ \eqref{eq:64a} arising from the SP$_3$ equations,
$$
\mu^+ = \frac{1}{\lambda^+} = 0.339981,
$$
$$
\mu^-=\frac{1}{\lambda^-} = 0.861137
$$
are the two positive values of $\mu$ in the S$_4$ Gauss-Legendre quadrature set.

Introducing Eqs.\ \eqref{eq:64} into Eq.\ \eqref{eq:62a}, we get
$$
a^+ = \frac{14}{35-9(2.941340)^2},
$$
$$
a^- = \frac{14}{35-9(1.161256)^2},
$$
or
\begin{subequations}\label{eq:65}
\begin{align}
a^+ = -0.326619,\label{eq:65a}\\
a^-=0.612334.\label{eq:65b}
\end{align}
\end{subequations}

Thus, we have found two solutions of Eqs.\ \eqref{eq:59} of the form defined by Eqs.\ \eqref{eq:61}: one for $\lambda^+$ and $a^+$ defined by Eqs.\ \eqref{eq:64a} and \eqref{eq:65a}; the other for $\lambda^-$ and $a^-$ defined by Eqs.\ \eqref{eq:64b} and \eqref{eq:65b}. The general solution is a linear combination of these two solutions, e.g.
\begin{subequations}\label{eq:66}
\begin{align}
G_0(r) = \Sigma_tA^+\left(\frac{e^{-\Sigma_t\lambda^+ r}}{4\pi r}\right)+\Sigma_tA^-\left(\frac{e^{-\Sigma_t\lambda^- r}}{4\pi r}\right),\label{eq:66a}\\
G_2(r) = \Sigma_tA^+a^+\left(\frac{e^{-\Sigma_t\lambda^+ r}}{4\pi r}\right)+\Sigma_tA^-a^-\left(\frac{e^{-\Sigma_t\lambda^- r}}{4\pi r}\right),
\end{align}
\end{subequations}
where the constants $A^+$ and $A^-$ will be determined by Eqs.\ \eqref{eq:60}. Inserting Eqs.\ \eqref{eq:66} into \eqref{eq:60}, we obtain
\begin{subequations}
\begin{align}
A^+a^++A^-a^- = -\frac{14}{9},\\
A^++A^- = \frac{55}{9}.
\end{align}
\end{subequations}
Solving these equations for $A^+$ and $A^-$, we obtain
\begin{subequations}\label{eq:68}
\begin{align}
A^+ = 5.642025.\\
A^- = 0.469086.
\end{align}
\end{subequations}

Thus, the Green's function for the scalar flux $G_0(r)$ is given by Eq.\ \eqref{eq:66a}, with $\lambda^{\pm}$ defined by Eqs.\ \eqref{eq:64} and $A^{\pm}$ by Eqs.\ \eqref{eq:68}: 
\begin{subequations}\label{eq:69}
\begin{align}
G_0(r) = \frac{\Sigma_t}{4\pi r}\left[A^+e^{-\Sigma_t\lambda^+ r}+A^-e^{-\Sigma_t\lambda^-r}\right],
\end{align}
where
\begin{align}
\lambda^+ = 2.941340,\\
\lambda^- = 1.161256,
\end{align}
and
\begin{align}
A^+ = 5.642025,\\
A^-=0.469086.
\end{align}
\end{subequations}
Equation \eqref{eq:56a} now gives:
\begin{equation*}
\begin{split}
\Sigma_t\phi_0(\ux)&=\int\int\int\Sigma_tG_0(|\ux-\ux'|)S(\ux')dV' \\
&= \int\int\int\frac{\Sigma_t^2|\ux-\ux'|\left[A^+e^{-\Sigma_t\lambda^+|\ux-\ux'|}+A^-e^{-\Sigma_t\lambda^-|\ux-\ux'|}\right]S(\ux')}{4\pi|\ux-\ux'|^2}dV',
\end{split}
\end{equation*}
and this agrees with Eq.\ \eqref{eq:20a} if we define:
\begin{equation}
p(s) = \Sigma_t^2s\left(A^+e^{-\Sigma_t\lambda^+s}+A^-e^{-\Sigma_t\lambda^-s}\right), \hspace{1cm} 0\leq s<\infty.
\end{equation}
To confirm that this legitimately defines a distribution function, we can easily calculate:
$$
\int_0^\infty\Sigma_t^2s\left(A^+e^{-\Sigma_t\lambda^+s}+A^-e^{-\Sigma_t\lambda^-s}\right)ds = \frac{A^+}{(\lambda^+)^2}+\frac{A^-}{(\lambda^-)^2} = 1.
$$
Thus,
$$
\int_0^\infty p(s)ds=1,
$$
as required.

Next, one can easily obtain
\begin{equation*}
\begin{split}
\int_s^\infty p(s')ds' &= \int_s^\infty\Sigma_t^2s'\left(A^+e^{-\Sigma_t\lambda^+s'}+A^-e^{-\Sigma_t\lambda^-s'}\right)ds' \\
&= A^+\left(\frac{1+\Sigma_t\lambda^+s}{(\lambda^+)^2}\right)e^{-\Sigma_t\lambda^+s}+A^-\left(\frac{1+\Sigma_t\lambda^-s}{(\lambda^-)^2}\right)e^{-\Sigma_t\lambda^-s}.
\end{split}
\end{equation*}
Therefore,
\begin{equation}
\label{eq:71}
\Sigma_t(s) = \frac{p(s)}{\int_s^\infty p(s')ds'} = \frac{A^+(\Sigma_t^2 s)e^{-\Sigma_t\lambda^+s}+A^-(\Sigma_t^2 s)e^{-\Sigma_t\lambda^-s}}{A^+\left(\frac{1+\Sigma_t\lambda^+s}{(\lambda^+)^2}\right)e^{-\Sigma_t\lambda^+s} + A^-\left(\frac{1+\Sigma_t\lambda^-s}{(\lambda^-)^2}\right)e^{-\Sigma_t\lambda^-s}}.
\end{equation}
For $\Sigma_ts\gg 1$,
$$
e^{-\Sigma_t \lambda^+s}\ll e^{-\Sigma_t\lambda^-s},
$$
and Eq.\ \eqref{eq:71} reduces to
\begin{equation}
\Sigma_t(s) = \frac{\Sigma_t^2s}{1+\Sigma_t\lambda^-s}(\lambda^-)^2\approx\Sigma_t\lambda^-\approx 1.161256\Sigma_t \hspace{1cm} (s\rightarrow \infty).
\end{equation}
This result is more accurate than the ``diffusion" value of $\sqrt{3}\Sigma_t=1.732051\Sigma_t$.

The SP$_3$ mean free path is:
\begin{equation}
\begin{split}
\bar s = \int_0^\infty sp(s)ds &= \int_0^\infty\Sigma_t^2s^2\left(A^+e^{-\Sigma_t\lambda^+s} + A^-e^{-\Sigma_t\lambda^-s}\right)ds\\
& = \frac{1}{\Sigma_t}\left(\frac{2A^+}{(\lambda^+)^3}+\frac{2A^-}{(\lambda^-)^3}\right)\\
& = 
\frac{1.042533}{\Sigma_t}.
\end{split}
\end{equation}
This result is, of course, much closer to the ``correct" value of $\Sigma_t^{-1}$ than either the diffusion result [Eq.\ \eqref{eq:28}] or the SP$_2$ result [Eq.\ \eqref{eq:51}].

Finally, the distance-to-collision $s$ can be sampled by the formula
\begin{equation}
\begin{split}
\xi &= \int_0^s p(r)dr = \int_0^s\left[\Sigma_t^2r\left(A^+e^{-\Sigma_t\lambda^+r} + A^-e^{-\Sigma_t\lambda^-r}\right)\right]dr\\
& = \frac{A^+}{(\lambda^+)^2}\left[1-(1+\Sigma_t\lambda^+s)e^{-\Sigma_t\lambda^+s}\right] +
\frac{A^-}{(\lambda^-)^2}\left[1-(1+\Sigma_t\lambda^-s)e^{-\Sigma_t\lambda^-s}\right] \\
& = F(\Sigma_ts), \hspace{1cm} 0<s<\infty.
\end{split}
\end{equation}
The function $F$ can be tabulated to efficiently give
\begin{equation}
s = \frac{1}{\Sigma_t}F^{-1}(\xi).
\end{equation}
Again, the functions $\Sigma_t(s)$, $p(s)$, and $\xi(s)$ are shown in Figures \ref{fig:Sigmat}-\ref{fig:xi}.

\section{Discussion}\label{sec:7}
%

In this paper we have shown that for an infinite homogeneous medium, three diffusion-based approximations to the standard steady-state linear Boltzmann equation (classic diffusion, $SP_2$, and $SP_3$) can each be represented \emph{exactly} by a non-classical transport equation with a non-constant  $\Sigma_t(s)$. (For the standard Boltzmann equation, $\Sigma_t(s)=\Sigma_t=\text{constant}$, independent of $s$.) The practical value of the approximate diffusion-based methods is that they are traditionally formulated without use of the angular variable $\uomega$, making them much less expensive to simulate than the original Boltzmann equation.
For each diffusion approximation, we derived an explicit expression for the path-length distribution $p(s)$ and showed that as one progresses from diffusion to $SP_2$ to $SP_3$, the corresponding scattering cross section $\Sigma_t(s)$ increasingly better-approximates the constant $\Sigma_t$ of the Boltzmann equation (see Figure 1). This is because the classical exponential distribution is approximated better and better (see Figure 2). As a result, we have seen that the mean free path is approximated with increasing accuracy. As a side note, we remark that the second moment of the path length distribution $\int_0^\infty s^2p(s)ds$ for all diffusion approximations gives the exact transport value $\frac{2}{\Sigma_t^2}$. However, we do not see a systematic reason why that should be the case. 

These results give theoretical insight into the properties of the approximate methods. The results also make it possible -- in principle -- to consistently simulate diffusion, $SP_2$, and $SP_3$ problems using a Monte Carlo method in which the distance-to-collision is determined by a non-exponential distribution function. However, before this can be done for realistic problems, the theory in this paper must be generalized in two ways:

	First, the theory must be extended to heterogeneous media, in such a way that the interface conditions for the non-classical Boltzmann equation at material interfaces are consistent with the interface conditions used for the relevant diffusion approximations. The ``natural" interface condition for the non-classical Boltzmann equation would seem to be that for each $\ux$ and $\uomega$  the non-classical angular flux should be a continuous function of $s$. Presumably, this (or some other) condition is consistent with the standard approximate interface conditions. 

	Second, the theory in this paper must be extended to finite media. For such problems, boundary conditions for the non-classical  angular flux on the outer boundary of the system must be formulated in a way that is consistent with standard outer boundary conditions to the relevant diffusion approximation. Specifically, what should the assigned value of $s$ be for particles that enter the system from the exterior? The choice $s=0$ is intuitively appealing but is not necessarily correct. 

	This work must be done in order for the representation of the diffusion-based approximations to the Boltzmann equation by non-classical Boltzmann equations to become ``complete." When this happens, it will be possible to interpret these approximations as being fully equivalent to non-classical transport processes, and to employ Monte Carlo methods to directly simulate them. However, these remaining tasks must be left for future work.

\end{document}